\documentclass[journal=apchd5,manuscript=letter]{achemso}

\usepackage{chemformula} 
\usepackage[T1]{fontenc} 

\author{Alison E. Rugar}
\altaffiliation{These authors contributed equally to this work.}
\affiliation{E. L. Ginzton Laboratory, Stanford University, Stanford, CA 94305, USA}
\email{arugar@stanford.edu}
\author{Constantin Dory}
\altaffiliation{These authors contributed equally to this work.}
\affiliation{E. L. Ginzton Laboratory, Stanford University, Stanford, CA 94305, USA}
\author{Shahriar Aghaeimeibodi}
\altaffiliation{These authors contributed equally to this work.}
\affiliation{E. L. Ginzton Laboratory, Stanford University, Stanford, CA 94305, USA}
\author{Haiyu Lu}
\affiliation{Department of Physics, Stanford University, Stanford, California 94305, United States}
\alsoaffiliation{Geballe Laboratory for Advanced Materials, Stanford University, Stanford, California 94305, United States}
\author{Shuo Sun}
\affiliation{E. L. Ginzton Laboratory, Stanford University, Stanford, CA 94305, USA}
\author{Sattwik Deb Mishra}
\affiliation{E. L. Ginzton Laboratory, Stanford University, Stanford, CA 94305, USA}
\author{Zhi-Xun Shen}
\affiliation{Department of Physics, Stanford University, Stanford, California 94305, United States}
\alsoaffiliation{Department of Applied Physics, Stanford University, Stanford, California 94305, United States}
\alsoaffiliation{Geballe Laboratory for Advanced Materials, Stanford University, Stanford, California 94305, United States}
\alsoaffiliation{Stanford Institute for Materials and Energy Sciences, SLAC National Accelerator Laboratory, Menlo Park, California 94025, United States}
\author{Nicholas A. Melosh}
\affiliation{Stanford Institute for Materials and Energy Sciences, SLAC National Accelerator Laboratory, Menlo Park, California 94025, United States}
\alsoaffiliation{Department of Materials Science and Engineering, Stanford University, Stanford, California 94305, United States}
\alsoaffiliation{Geballe Laboratory for Advanced Materials, Stanford University, Stanford, California 94305, United States}
\author{Jelena Vu\v{c}kovi\'c}
\affiliation{E. L. Ginzton Laboratory, Stanford University, Stanford, CA 94305, USA}

\title[An \textsf{achemso} demo]
  {Narrow-linewidth tin-vacancy centers in a diamond waveguide}


\begin{document}


\begin{abstract}
Integrating solid-state quantum emitters with photonic circuits is essential for realizing large-scale quantum photonic processors. Negatively charged tin-vacancy (SnV$^-$) centers in diamond have emerged as promising candidates for quantum emitters because of their excellent optical and spin properties including narrow-linewidth emission and long spin coherence times. SnV$^-$ centers need to be incorporated in optical waveguides for efficient on-chip routing of the photons they generate. However, such integration has yet to be realized. In this Letter, we demonstrate the coupling of SnV$^-$ centers to a nanophotonic waveguide. 
We realize this device by leveraging our recently developed shallow ion implantation and growth method for generation of high-quality SnV$^-$ centers and the advanced quasi-isotropic diamond fabrication technique.
We confirm the compatibility and robustness of these techniques through successful coupling of narrow-linewidth SnV$^-$ centers (as narrow as $36\pm2$~MHz) to the diamond waveguide. Furthermore, we investigate the stability of waveguide-coupled SnV$^-$ centers under resonant excitation. Our results are an important step toward SnV$^-$-based on-chip spin-photon interfaces, single-photon nonlinearity, and photon-mediated spin interactions. 
\end{abstract}

\section{Introduction}
The realization of large-scale optical quantum information processing requires the integration of quantum emitters into a photonic infrastructure that guides and manipulates light\cite{Kimble_Quantum_Internet}. In recent years, several color centers in solids have been explored as potential optically interfaced qubit candidates\cite{Awschalom2018}. In particular, group-IV color centers in diamond, such as the negatively charged silicon-\cite{NeuSiV_Iridium,Hepp_StructureSiV_PRL_2014,Mueller_SiVSpins_NatComm_2014}, germanium-\cite{IwasakiGeV2015,PalyanovGeV2015, Ekimov_GeV_2015}, tin-\cite{IwasakiSnV,TchernijSnV,Ekimov_SnV_HPHT2018,Rugar_SnV_PRB2018,Trusheim_SnV_PRL2020}, and lead-vacancy\cite{EnglundPbV,TchernijPbV} (SiV$^-$, GeV$^-$, SnV$^-$, and PbV$^-$ respectively) centers have garnered much attention because of their favorable optical properties and robustness to electric-field fluctuations to first order, due in large part to their inversion-symmetric structure\cite{EvansSiVPRApplied2016,AharonovichTrusheimReview2019}. While remarkable progress has been made in SiV$^-$ and GeV$^-$ quantum photonics\cite{LukinSiVQuantumRegister2019_shorter,Bhaskar_MemoryEnhanced_SiV_Nature_2020,Bhaskar_GeV_waveguide_PRL_2017,AharonovichGeV2018},
it is important to extend those efforts to other color centers with comparable or superior properties in order to find the best possible qubit candidate.
The SnV$^-$ center is of specific interest because it is expected to have long spin coherence times at temperatures above $1$~K\cite{GaliPRX,Trusheim_SnV_PRL2020}, unlike the negatively charged SiV$^-$ and GeV$^-$ centers which require dilution refrigerator temperatures\cite{AharonovichTrusheimReview2019}. However, despite its exciting potential as an optically interfaced qubit candidate, the SnV$^-$ center has not yet been incorporated into any nanophotonic devices that can be scaled to eventually form integrated quantum photonic circuits.

In this paper, we present the first demonstration of SnV$^-$ centers coupled to a waveguide, a fundamental building block for waveguide quantum electrodynamics\cite{Bhaskar_GeV_waveguide_PRL_2017, SipahigilScience2016, Grim_NatureMaterials_2019} and large-scale photonic circuitry. SnV$^-$ centers are generated via the shallow ion implantation and growth (SIIG) method\cite{Rugar_SnV_SIIG_NanoLett2020}, and the waveguide is fabricated using a quasi-isotropic etch technique\cite{Mouradian1DPhC2017,Wan2DPhC2018,MitchellMicrodisksAPLPhotonics,DoryOptimizedDiamondPhotonics}. The measured linewidth of a resulting SnV$^-$ center coupled to a waveguide in our sample is on par with those measured in previous reports of lifetime-limited SnV$^-$ centers\cite{Trusheim_SnV_PRL2020,Goerlitz_SnV_NJP2020}. The previously characterized SnV$^-$ centers were either in bulk diamond\cite{Goerlitz_SnV_NJP2020} or micropillars\cite{Trusheim_SnV_PRL2020} and were generated via high-energy implantation and annealing in vacuum\cite{Trusheim_SnV_PRL2020} or high-pressure, high-temperature (HPHT) environments\cite{Goerlitz_SnV_NJP2020}.
Our results demonstrate that the SIIG method and fabrication based on a quasi-isotropic undercut can be combined to form a platform for quantum photonics experiments with high-quality SnV$^-$ centers. 
Having narrow-linewidth SnV$^-$ centers in nanophotonic structures in diamond paves the way to more advanced on-chip quantum optics experiments performed at higher temperatures than previously allowed by the SiV$^-$ center.
We also provide detailed characterization and statistical analysis of the stability of the waveguide-coupled emitters.

\section{Sample preparation}
\begin{figure}[H]
\includegraphics[width=0.8\textwidth,]{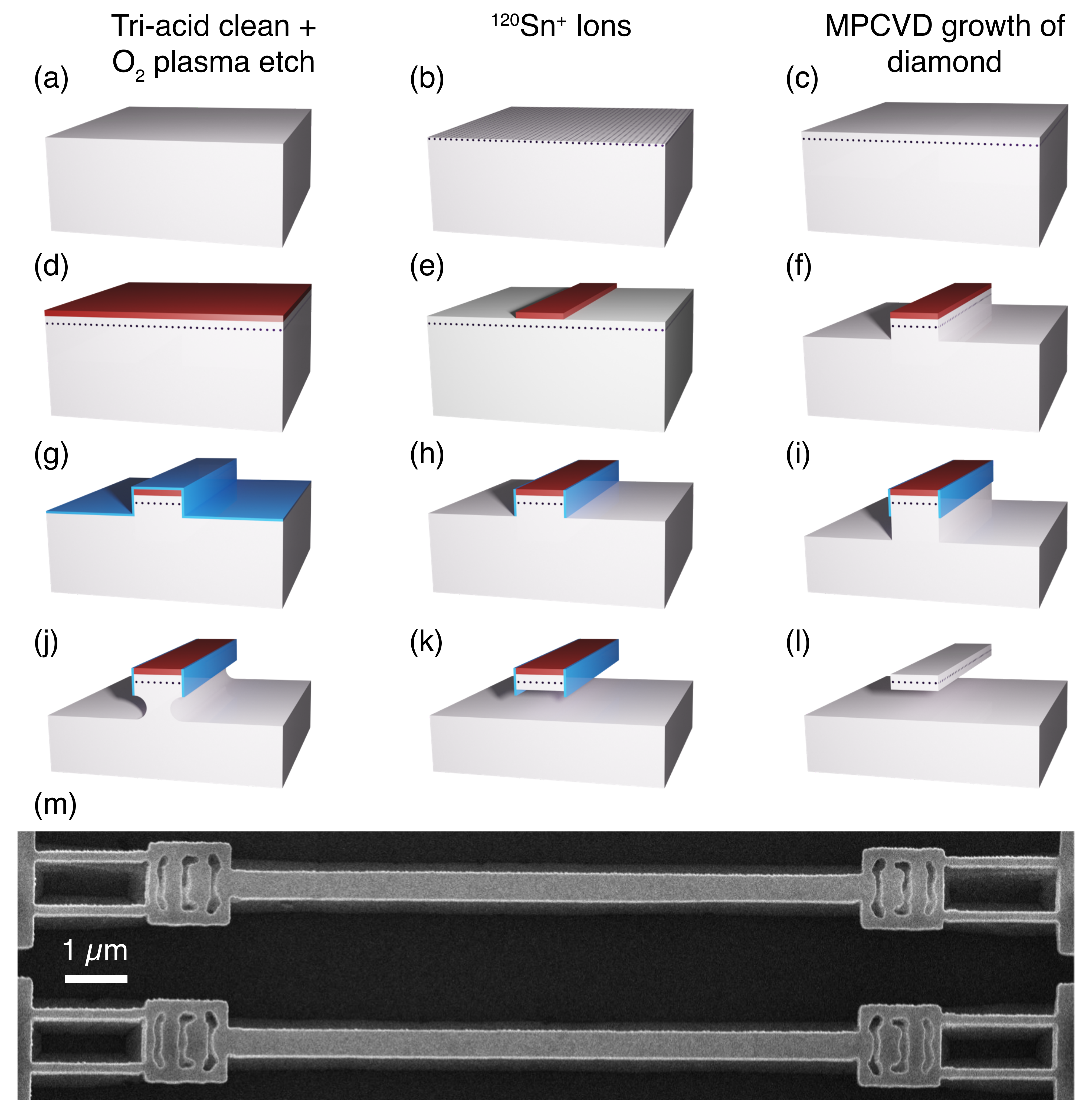}
\caption{Sample fabrication. (a)--(l) Schematics of sample fabrication. (a) Tri-acid clean and etch of electronic grade diamond with O$_2$ plasma. (b) Implantation of $^{120}$Sn$^+$ ions at 1~keV with a dose of $1.6\times10^{10}$~cm$^{-2}$. (c) H$_2$ plasma clean and MPCVD growth of 90~nm of diamond. (d) PECVD growth of 200 nm of Si$_x$N$_y$ (red) as an etch mask. (e) Electron-beam lithography and SF$_6$, CH$_4$, and N$_2$ RIE to transfer pattern to the Si$_x$N$_y$ etch mask. (f) Anisotropic O$_2$ dry etch of diamond. (g) Atomic layer deposition of 30~nm of Al$_2$O$_3$ (blue). (h) Anisotropic dry etch of Al$_2$O$_3$. (i) Second anisotropic O$_2$ dry etch of diamond. (j) Quasi-isotropic etch of diamond along the \{110\} facet allows the undercut to form. (k) Continuation of quasi-isotropic etch until desired thickness is attained. Higher etch rate along \{110\} compared to \{001\} leads to a flat bottom surface. (l) Removal of Si$_x$N$_y$ and Al$_2$O$_3$ masks. (m) Typical SEM image of resulting devices.}
\label{sampleprep_fig}
\end{figure}

An electronic grade diamond chip from ElementSix is first cleaned in a boiling tri-acid solution consisting of nitric, sulfuric, and perchloric acids in equal part. The top $\sim500$~nm of the diamond chip is then removed via an oxygen (O$_2$) plasma etch (Figure~1(a)). The subsequent color center generation and nanofabrication steps are detailed below and illustrated schematically in Figure~\ref{sampleprep_fig}.

SnV$^-$ centers are generated on the chip via the shallow ion implantation and growth (SIIG) method\cite{Rugar_SnV_SIIG_NanoLett2020}. The prepared chip is implanted by CuttingEdge Ions with $^{120}$Sn$^{+}$ ions at 1~keV and a dose of $1.6\times10^{10}$~cm$^{-2}$ (Figure~1(b)). The implanted chip is then cleaned with a hydrogen (H$_2$) plasma immediately before 90~nm of diamond is grown via microwave-plasma chemical vapor deposition (MPCVD, Seki Diamond Systems SDS 5010) 
as detailed in Ref. \citenum{Rugar_SnV_SIIG_NanoLett2020} and shown in Figure~1(c).

With SnV$^-$ centers embedded 90~nm below the diamond surface, we fabricate suspended bare waveguides. 
We first grow 200~nm of Si$_x$N$_y$ via plasma-enhanced chemical vapor deposition (PECVD) to serve as an etch mask (Figure~\ref{sampleprep_fig}(d)). We define the structures in hydrogen silsesquioxane FOx-16 via electron-beam lithography and transfer the pattern into the Si$_x$N$_y$ with a SF$_6$, CH$_4$, and N$_2$ reactive ion etch (RIE) (Figure~\ref{sampleprep_fig}(e)). With the Si$_x$N$_y$ etch mask patterned, we then etch the diamond with an anisotropic O$_2$ RIE (Figure~\ref{sampleprep_fig}(f)). Next, 30~nm of Al$_2$O$_3$ are grown via atomic layer deposition (Figure~\ref{sampleprep_fig}(g)). The horizontal planes of the Al$_2$O$_3$ layer are removed with a Cl$_2$, BCl$_2$, and N$_2$ RIE (Figure~\ref{sampleprep_fig}(h)). A second anisotropic O$_2$ RIE is performed to expose part of the vertical sidewalls of the diamond structure (Figure~\ref{sampleprep_fig}(i)). The quasi-isotropic etch\cite{Mouradian1DPhC2017,Wan2DPhC2018,MitchellMicrodisksAPLPhotonics,DoryOptimizedDiamondPhotonics} step can now be performed to undercut the structures (Figure~\ref{sampleprep_fig}(j)). To do so, the forward bias of the O$_2$ RIE is turned off, and the etch is performed at high temperature with high inductively coupled plasma power\cite{Mouradian1DPhC2017,Wan2DPhC2018,MitchellMicrodisksAPLPhotonics,DoryOptimizedDiamondPhotonics}. The resulting etch progresses preferentially along the \{110\} planes\cite{Mouradian1DPhC2017}. When the nanobeam waveguides have been released (Figure~\ref{sampleprep_fig}(k)) and have been etched to the desired thickness, as validated by high-voltage scanning electron microscopy (SEM) imaging, the sample is soaked in hydrofluoric acid to remove the Si$_x$N$_y$ and Al$_2$O$_3$ etch masks (Figure~\ref{sampleprep_fig}(l)).

The resulting waveguides are 400~nm wide, 10~$\mu$m long, and 280~nm thick. On either end of the waveguides is an inverse-designed\cite{Molesky_InvDesReview_NaturePhton_2018,DoryOptimizedDiamondPhotonics} vertical coupler (VC) to couple incident light from free space into the waveguide and vice versa. A SEM image of a typical resulting nanostructure is shown in Figure~\ref{sampleprep_fig}(m). We successfully fabricated more than 1000 waveguide structures across our 2~$\times$~2 mm$^2$ chip.

\section{Results and discussion}
\subsection{Photoluminescence map}
\begin{figure}[h]
\includegraphics[width=0.5\textwidth]{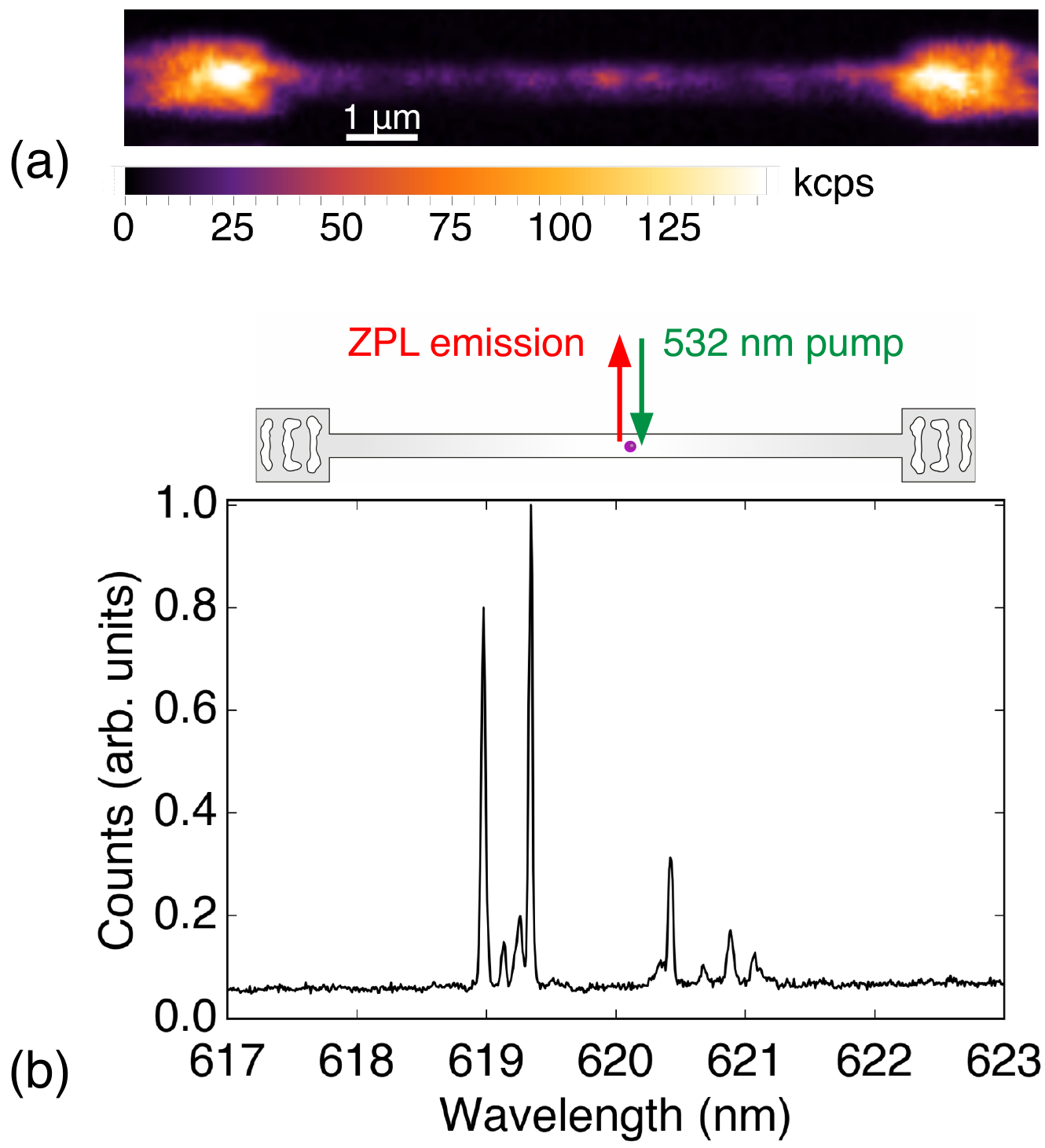}
\caption{PL of SnV$^-$ centers in a waveguide. (a) PL map of a waveguide on the chip containing SnV$^-$ centers. (b) PL spectrum collected on a waveguide with excitation and collection (MM fiber) aligned on the same spot. The configuration is shown schematically above the plot.}
\label{PL_fig}
\end{figure}

To characterize the fabricated devices, we use two home-built cryogenic setups. First, we acquire two-dimensional photoluminescence (PL) maps of potential structures using a scanning confocal microscope operating at 5 K. In this setup, an objective lens inside the cryostat (Montana Instruments) focuses a continuous-wave laser at 532 nm on the sample to excite the emitters non-resonantly. The emitted photons are collected using the same objective lens and a single-mode fiber. We spectrally filter the collected light using a 568-nm long-pass filter and send the filtered signal to a single-photon counting module. A typical resulting PL map of a nanobeam waveguide on this sample is shown in Figure~\ref{PL_fig}(a). The high counts and distinct bright spots apparent on the device in the PL map indicate the existence of SnV$^-$ centers in the nanobeam. PL spectroscopy is next used to confirm the presence of SnV$^-$ centers in the devices.

\subsection{Photoluminescence excitation}
To characterize the linewidth of the emitters, we move to another cryostat (attoDRY2100) that operates at 1.7 K. We first collect a PL spectrum, shown in Figure~\ref{PL_fig}(b) by focusing a 532-nm laser on a spot in the middle of the waveguide, collecting the emitted light from the same spot with a multimode fiber, and sending the collected light to a spectrometer. This spectrum displays multiple distinct zero-phonon lines clustered around the expected wavelengths of the C and D transitions\cite{IwasakiSnV} of the SnV$^-$ center, indicating that multiple SnV$^-$ centers are present in the nanobeam waveguide. In the Supporting Information\cite{supportinginfo}, we inspect several other waveguides and confirm the presence of SnV$^-$ centers in all of the studied structures.

We perform photoluminescence excitation (PLE) on the C transition of a color center in the same waveguide. We scan the wavelength of the MSquared External Mixing Module around the resonant wavelength of the color center of interest. Counting photons emitted into the phonon sideband of the SnV$^-$ center, we observe a peak in emission as the laser crosses resonance with the color center and thus can infer the linewidth of the color center. We perform PLE in two configurations: top-down and VC-to-VC. In the former, illustrated schematically in Figure~\ref{VC2VC_PLE_fig}(a), the excitation laser is coupled out of a single-mode polarization maintaining fiber and is focused on a spot on the waveguide. The collection fiber is aligned to the same spot on the waveguide. In the VC-to-VC configuration, illustrated schematically in Figure~\ref{VC2VC_PLE_fig}(b), the excitation spot is focused on one VC.
The emission is collected from the opposite VC. In both configurations, the emitted light is passed through a 638-nm long-pass filter and a 700-nm short-pass filter before being collected by a multimode fiber. 

\subsubsection{Narrow-linewidth SnV$^-$ centers in waveguides}
\begin{figure}[t]
\includegraphics[width=0.7\textwidth]{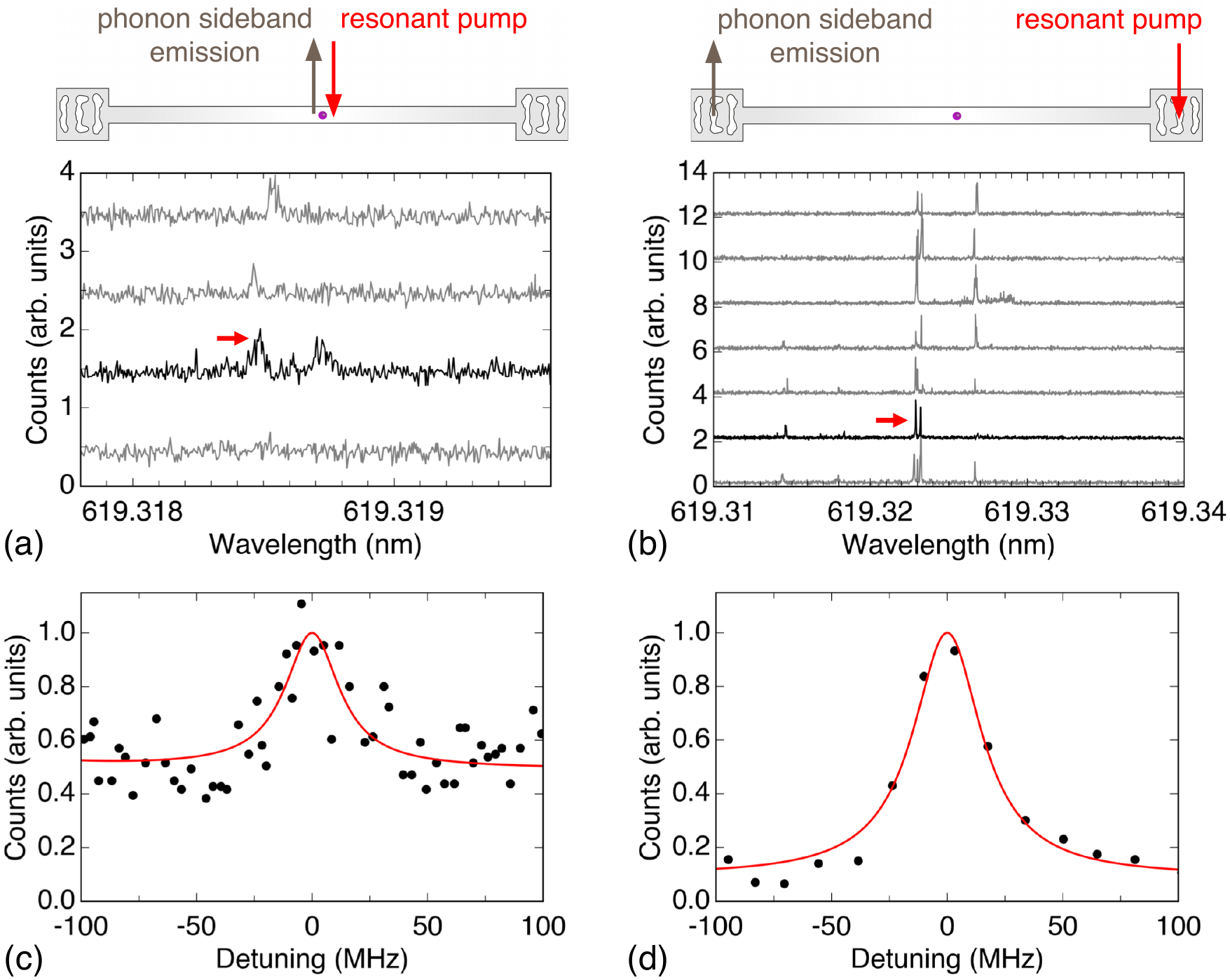}
\caption{PLE of a SnV$^-$ center in a waveguide. (a) PLE scans in the top-down configuration, as shown schematically above the plot. Repeated scans are offset vertically for clarity. (b) PLE scans in the VC-to-VC configuration, as shown schematically above the plot. Repeated scans are offset vertically for clarity. The red arrows in (a) and (b) indicate a peak that we investigate more closely in the next panels. (c) Close-up view of the peak indicated by the red arrow in (a). A Lorentzian fit (red curve) to the data (black dots) reveals a linewidth of $29\pm5$~MHz. (d) Close-up view of the peak indicated by the red arrow in (b). A Lorentzian fit (red curve) to the data (black dots) reveals a linewidth of $36\pm2$~MHz.}
\label{VC2VC_PLE_fig}
\end{figure}

Figures~\ref{VC2VC_PLE_fig}(a) and (b) display the resulting PLE data for the top-down and VC-to-VC configurations respectively. 
The different passes of the laser through resonance are offset in the plot for clarity, with the lowest scan being the first. The many peaks in Figure~\ref{VC2VC_PLE_fig}(b) may indicate that several emitters are coupled to the waveguide, as expected from the PL shown in Figure~\ref{PL_fig}.

From the data in Figures~\ref{VC2VC_PLE_fig}(a) and (b), we fit a Lorentzian to one peak for each configuration, indicated by a red arrow. The Lorentzian fits to these data, shown in Figures~\ref{VC2VC_PLE_fig}(c) and (d), reveal linewidths of $29\pm5$~MHz and $36\pm2$~MHz respectively. Notably, both of these linewidths are comparable to previously reported lifetime-limited linewidths\cite{Trusheim_SnV_PRL2020, Goerlitz_SnV_NJP2020} despite the several fabrication steps performed on this sample. 
The narrow linewidths of the SnV$^-$ centers presented in Figure~\ref{VC2VC_PLE_fig} are reproducible in other SnV$^-$ centers. PLE and linewidth data of several additional SnV$^-$ centers are provided in the Supporting Information \cite{supportinginfo}. The average linewidth of these other SnV$^-$ centers is $36 \pm 3$~MHz.
The ability to produce narrow-linewidth SnV$^-$ centers in suspended waveguides will enable the future fabrication of large-scale photonic circuitry and quantum photonics experiments.

\subsubsection{Blinking and spectral diffusion}
In both configurations, Figures~\ref{VC2VC_PLE_fig}(a) and (b), multiple peaks are apparent in several scans but not all appear in the same scans. The inconsistent appearances of some peaks, known as blinking, may indicate instabilities in the charge states of those emitters\cite{Goerlitz_SnV_NJP2020}. We also observe that the lines vary slightly in frequency between PLE scans, or display spectral diffusion.
The fact that these instabilities are observed in both configurations confirms that this is a color-center-related phenomenon rather than a waveguide-related one.
Similar observations have been reported in Ref.~\citenum{Goerlitz_SnV_NJP2020} where SnV$^-$ centers were generated via high-energy ion implantation and HPHT annealing.

\begin{figure}[t]
\includegraphics[width=0.8\textwidth]{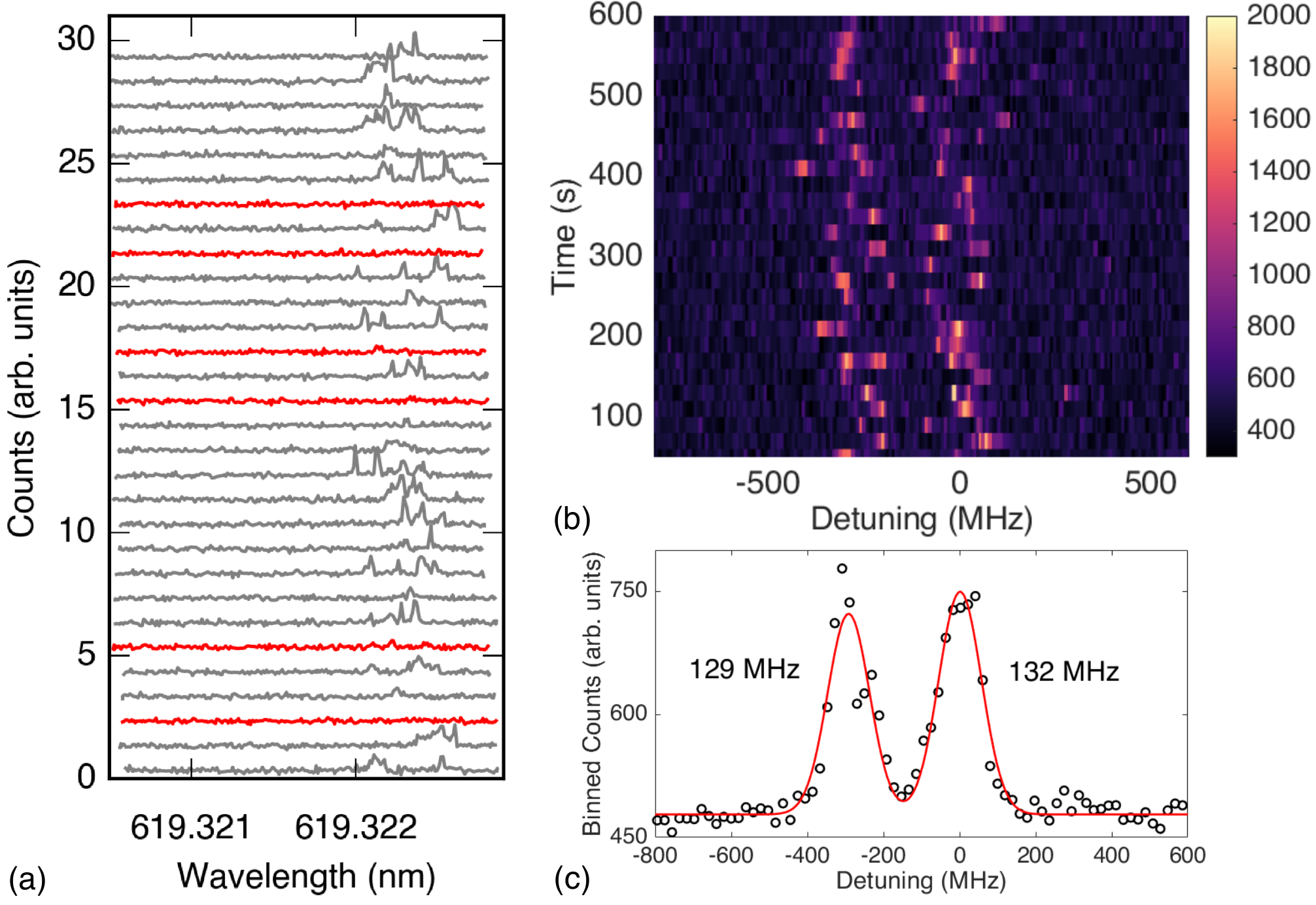}
\caption{Blinking and spectral diffusion. (a) Consecutive PLE scans acquired for ten minutes, showing blinking. Scans are offset vertically for clarity. Scans in which blinking occurred are plotted in red. (b) Heat map of PLE accumulated over ten minutes at 50-nW excitation in the top-down configuration, for a different SnV$^-$ center. Legend is in units of counts per second. (c) Histogram generated from data presented in (b). Spectral diffusion is quantified based on a Gaussian fit (red curve) to the histogram.}
\label{diffusion_blinking_fig}
\end{figure}

In Figure~\ref{diffusion_blinking_fig}(a), we highlight in red the PLE scans in which the SnV$^-$ center under study blinked. 
In a survey of SnV$^-$ centers in waveguides, we found that seven of the nine SnV$^-$ centers that we studied blinked at least once during ten minutes of consecutive PLE scans. Additional details on the quantitative definition of blinking and the dependence of blinking on excitation power can be found in the Supporting Information \cite{supportinginfo}.

Figures~\ref{diffusion_blinking_fig}(b) and (c) show an example of the analysis of the spectral diffusion of SnV$^-$ centers in a waveguide. Consecutive PLE scans are acquired for ten minutes (Figure~\ref{diffusion_blinking_fig}(b)). The data are then binned in frequency to form a histogram of counts at different frequencies, as shown in Figure~\ref{diffusion_blinking_fig}(c). 
Because of the random nature of the spectral diffusion, we fit a Gaussian to the histogram. We quantify the spectral diffusion by calculating the full width at half maximum of the fit.
In Figures~\ref{diffusion_blinking_fig}(b) and (c), the distribution is bimodal with full widths at half maximum of about 130~MHz. A statistical study of the spectral diffusion of several SnV$^-$ centers and data on the power dependence of spectral diffusion are provided in the Supporting Information \cite{supportinginfo}.

\section{Conclusion and outlook}
In this work we have integrated narrow-linewidth SnV$^-$ centers in diamond nanobeam waveguides. These results indicate that the combination of the SIIG method\cite{Rugar_SnV_SIIG_NanoLett2020} and the quasi-isotropic undercut\cite{Mouradian1DPhC2017,Wan2DPhC2018,MitchellMicrodisksAPLPhotonics,DoryOptimizedDiamondPhotonics} forms a promising platform for future quantum photonics with the SnV$^-$ center. For example, carefully optimizing the coupling efficiency of the narrow-linewidth emitter to the waveguide mode enables a broadband interface for spin-controlled photon switching\cite{javadi2018spin} and few-photon nonlinearity\cite{Thy_lifetime_limited_QD_Nanolett_2018}. We have also investigated the spectral diffusion and blinking of the waveguide-coupled SnV$^-$ centers. Electromechanically controlled waveguides may increase the stability of emitters by reducing the charge and strain related noise\cite{SiV_PRX2019}. However, further experimental efforts are required to understand and to mitigate the source of instability in the emission spectra of the SnV$^-$ centers.

To go beyond single-emitter experiments, deterministic positioning of the SnV$^-$ centers along the waveguide is essential. Fortunately, as previously demostrated in Ref.~\citenum{Rugar_SnV_SIIG_NanoLett2020}, we can achieve site-controlled generation of SnV$^-$ centers with an implantation mask defined with standard lithography. Deterministic positioning combined with small inhomogeneous broadening of SnV$^-$ centers will enable the coupling of multiple identical emitters to the same optical mode for superradiance and quantum few-body experiments\cite{SipahigilScience2016, Grim_NatureMaterials_2019,Kim_Superradiance_NanoLetters_2018}. 

Moreover, having high-quality SnV$^-$ centers in suspended nanophotonic structures in diamond enables the future integration of the color centers into cavities to enhance the spin-photon interaction. Ultimately, such devices will be beneficial for several quantum information applications and can operate at elevated temperatures compared to previous demonstrations with SiV$^-$ centers\cite{Bhaskar_MemoryEnhanced_SiV_Nature_2020}.

\begin{acknowledgement}
This work is financially supported by Army Research Office (ARO) (award no. W911NF-13-1-0309); National Science Foundation (NSF) RAISE TAQS (award no. 1838976); Air Force Office of Scientific Research (AFOSR) DURIP (award no. FA9550-16-1-0223); Department of Energy, Basic Energy Sciences (BES) - Materials Science and Engineering (award no. DE-SC0020115). SIMES work is supported by the DOE Office of Sciences, Division of Materials Science and Engineering; and SLAC LDRD. A.E.R. acknowledges support from the National Defense Science and Engineering Graduate (NDSEG) Fellowship Program, sponsored by the Air Force Research Laboratory (AFRL), the Office of Naval Research (ONR) and the Army Research Office (ARO). C.D. acknowledges support from the Andreas Bechtolsheim Stanford Graduate Fellowship and the Microsoft Research PhD Fellowship. S.A. acknowledges support from Bloch postdoctoral fellowship in quantum science and engineering from Stanford Q-FARM. Part of this work was performed at the Stanford Nanofabrication Facility (SNF) and the Stanford Nano Shared Facilities (SNSF), supported by the National Science Foundation under award ECCS-1542152.

\end{acknowledgement}

\providecommand{\latin}[1]{#1}
\makeatletter
\providecommand{\doi}
  {\begingroup\let\do\@makeother\dospecials
  \catcode`\{=1 \catcode`\}=2 \doi@aux}
\providecommand{\doi@aux}[1]{\endgroup\texttt{#1}}
\makeatother
\providecommand*\mcitethebibliography{\thebibliography}
\csname @ifundefined\endcsname{endmcitethebibliography}
  {\let\endmcitethebibliography\endthebibliography}{}

\end{document}